\documentstyle[epsf,graphicx]{mn}

\def\x2{$\chi^{2}$}

\def\ginga{{\sl Ginga} }
\def\asca{{\sl ASCA} }
\def\rosat{{\sl ROSAT} }
\def\exosat{{\sl EXOSAT} }

\def\sax{{\sl BeppoSAX} }
\def\xte{{\sl RXTE} }

\def\x2{$\chi^{2}$}

\newbox\grsign \setbox\grsign=\hbox{$>$} \newdimen\grdimen \grdimen=\ht\grsign
\newbox\simlessbox \newbox\simgreatbox \newbox\simpropbox
\setbox\simgreatbox=\hbox{\raise.5ex\hbox{$>$}\llap
     {\lower.5ex\hbox{$\sim$}}}\ht1=\grdimen\dp1=0pt
\setbox\simlessbox=\hbox{\raise.5ex\hbox{$<$}\llap
     {\lower.5ex\hbox{$\sim$}}}\ht2=\grdimen\dp2=0pt
\setbox\simpropbox=\hbox{\raise.5ex\hbox{$\propto$}\llap
     {\lower.5ex\hbox{$\sim$}}}\ht2=\grdimen\dp2=0pt

\def\cunits{$\rm cm^{-2}$}
\def\funits{$\rm erg~cm^{-2}~s^{-1}$}

\begin{document}

\title[{\sl RXTE} observations of Seyfert-2 galaxies]
{\bf  {\sl RXTE} observations of Seyfert-2 galaxies: evidence for spectral
 variability}
\author[I. Georgantopoulos \& I. E. Papadakis]
{\Large I. Georgantopoulos$^1$,  I. E. Papadakis $^2$  \\
$^1$ Institute of Astronomy \& Astrophysics, National Observatory of Athens, 
Lofos Koufou, Palaia Penteli, 15236, Athens, Greece \\
$^2$ Physics Department, University of Crete, 71003, Heraklion, Greece \\
}
\maketitle
\begin{abstract}
We present a series of \xte observations of the nearby 
 obscured Seyfert galaxies ESO103-G35, IC5063, NGC4507 and NGC7172.
  The period of monitoring ranges from seven days for NGC7172 
 up to about seven months for ESO103-G035.
 The spectra of all galaxies are well fit with 
 a highly obscured ($N_H>10^{23}$ \cunits) power-law and an Fe line at 6.4 keV.
 We find strong evidence for the presence of  a reflection component 
 in ESO103-G35 and NGC4507. The observed flux presents strong 
 variability on day timescales in all objects.
  Spectral variability is also detected in the sense that 
  the spectrum steepens with increasing flux 
   similar to the behaviour  
 witnessed in some Seyfert-1 galaxies.  
 
\end{abstract}

\newpage

\section{INTRODUCTION}

Recent X-ray missions have brought a rapid progress in our 
understanding of the X-ray properties of Seyfert galaxies  
(for a review see Mushotzky, Done \& Pounds 1993). 
The X-ray spectra of Seyfert 2 galaxies as observed by \ginga, \asca and 
recently {\it BeppoSAX}  have proved to be very complex 
 (Smith \& Done 1996; Turner et al. 1997). 
In broad terms most Seyfert 2 X-ray spectra can be well fitted by a power-law 
continuum (typically $\Gamma \sim 1.8$), plus an Fe-K emission line at 6.4 keV 
and a reflection component (e.g. Lightman 
\& White 1988, George \& Fabian 1991). This latter component,
 which suggests the presence of a large amount of 
 cold material in the vicinity of the nucleus, 
flattens the observed continuum and can dominate the spectrum 
above $\sim 10$ keV.  In most Seyfert 2s the above emission components 
are viewed through a large absorbing column density,  typically  
$N_H>10^{23}$ \cunits. This screen, which  
 suppresses the soft X-ray emission through the 
 process of photoelectric absorption in cool atomic and molecular gas, 
  is possibly associated with a large (pc scale) molecular torus.
  In the limit of very high column 
densities, Thomson scattering will also diminish the more penetrating 
hard X-ray emission.
 In some sources additional emission in
the form of a soft X-ray excess is observed below $\sim 3$ keV
probably as a result of scattering of the intrinsic power-law continuum by a
strongly photoionised medium (eg Griffiths et al. 1998 in the case of Markarian 3). 
 In order to observe such soft X-ray emission
it is clearly a requirement that the scattering medium should extend
  beyond the bounds of the obscuration 
of the molecular torus. 
The X-ray spectra of Seyfert-2 galaxies gave wide support to the 
 standard AGN unification model (Antonucci \& Miller 1985). 
 According to this paradigm, the nucleus (supermassive black hole, 
accretion disc and broad line region) has  basically the same structure 
in both type 1 and type 2 Seyferts, 
 but depending on the circumstances, can be hidden 
from view by the molecular torus (Krolik \& Begelman 1986). 
Specifically if the source is observed at a sufficiently high 
inclination angle,  and thus the line of sight 
intersects the torus, it would be classified as a Seyfert 
2, whereas for all other orientations it would be deemed to be a Seyfert 1.

In contrast to the recent progress in understanding the X-ray spectral 
characteristics of Seyfert 2 galaxies, our knowledge of their X-ray
variability properties remains  limited. According to the standard 
unification scenario, the hard X-ray continuum should vary with large 
amplitude in a similar way to that observed in Seyfert 1 galaxies 
(Mushotzky, Done \& Pounds 1993). 
 However, in Seyfert-2s a large fraction of the emission must come 
 from reprocessed radiation. 
 Since a large fraction of  the Fe-K line, the reflection
 emission and the soft excess are  likely to originate 
from regions of parsec scale-size, it follows that significantly less 
variability might be expected in Seyfert 2 objects, at least 
 in those parts of the spectrum where the reprocessing makes 
a substantial contribution to the 
overall flux. Time variability studies  can provide strong
 constraints on the geometry of the nucleus and the surounding region. 
For example, the time lag between the variation of the 
 power-law continuum and the Fe line flux or the reflection 
 component would give information on the location and the 
 size of the reflecting material. 
 In particular if a large amount of the Fe line 
 originates from the accretion disk, instead of the torus,
 we would expect variations of the line flux in timescales of days. 

Here, we present the results from several \xte archival observations   
 on four well known Seyfert-2 galaxies: ESO103-G35,  
 IC5063, NGC4507 and NGC7172. 
 In another paper (Georgantopoulos et al. 1999) we presented 
 monitoring observations of Markarian 3. 
The extended energy range 
of the {\sl RXTE} PCA  detectors and the large number of observations for
each object 
give us the opportunity to investigate in detail the properties of 
  the spectral components 
 present in these galaxies such as the intrinsic power-law and the reflection 
 component. 
Our main objective is to use the variability 
exhibited in the 3--24 keV band to place constraints on the geometry of 
the  nucleus and its circumnuclear matter.

\section{THE SAMPLE}

\subsection{ESO 103-G35} 

The HEAO-1 hard X-ray source 1H 1832-653 has been identified with the
galaxy ESO 103-G35 at a redshift z=0.013.  Optical spectroscopy revealed a
high excitation forbidden line spectrum with weak broad emission line
wings and therefore classified this object as a Seyfert-1.9 galaxy
(Phillips et al. 1979).  \exosat obtained the first X-ray spectrum showing
a power-law spectrum absorbed by a column of $N_H\sim 10^{23}$ \cunits
(Warwick, Pounds \& Turner 1988).  The \exosat observations also showed a
variation in the column density over a period of 90 days. Smith \& Done
(1996)  present \ginga observations detecting an Fe line around 6.4 keV
with an equivalent width of $\sim$ 350 eV. The spectrum of the source
could be characterised by a heavily absorbed power-law continuum of slope
$\Gamma \sim 1.8$ and $N_{H}\sim 2\times 10^{23}$ \cunits. Smith \& Done
(1996)  also found evidence for the presence of a reflection component.  
Turner et al. (1997) present the first \asca data for this object. They
resolve the Fe line into possibly three components (6.4, 6.68 and 6.96
keV).  Recently, Forster, Leighly \& Kay (1999) presented three \asca
observations separated by $\sim$ 2 years.  They find some marginal
evidence for a double Fe line in their first observation. Their best fit
$\Gamma$ and $N_{H}$ values agree with the \ginga results.  They also find
that the power-law continuum flux increased by a factor of two while the
Fe line equivalent width has decreased by about a similar amount. Two
observations of ESO103-G35 with {\it BeppoSAX} separated by a year, from
$\sim$2 to 60 keV (Akylas et al. 2000), show a variation of the
power-law flux by a factor of two, no variation of the Fe line 
flux while there is no strong evidence for the presence of a
reflection component.  Comparison of the historical data (Poletta et al.
1996)  from HEAO-1 up to ASCA show a variation of the flux of a factor of
about four.

\subsection{IC5063} IC5063 is an S0 galaxy (z=0.011)  presenting a typical
Seyfert-2 galaxy spectrum (Colina, Sparks \& Macchetto 1991). Scattered
broad $H_\alpha$ emission has been detected by Inglis et al. (1993). {\sl
HST} NICMOS observations show a very red unresolved point source (Kulkarni
et al. 1998).  IC5063 has a radio power at 1.4 GHz two orders of magnitude
greater than that of typical Seyfert galaxies  and
thus it can be classified as a Narrow-Line Radio galaxy as well (Ulvestad
\& Wilson 1984).  A \ginga observation showed a power-law photon index of
$\Gamma=1.5$ and $N_{H}\sim 2.5\times 10^{23}$ \cunits (Smith \& Done
1996). Turner et al. (1997) presented two \asca observations of IC5063.  
The spectrum is in good agreement with the \ginga observations.  IC5063
shows short-term X-ray flux variability at the 90 per cent confidence
level (Turner et al. 1997). However, the \ginga 
 and  the \asca observation obtained about four years later 
 give similar fluxes.

\subsection{NGC4507} NGC4507 is a nearby (z=0.012) barred spiral galaxy.
Optical spectra present high excitation, narrow emission lines classifying
it as a Seyfert-2 galaxy (Durret \& Bergeron 1986).  \ginga observations
showed a flat X-ray spectrum ($\Gamma \sim 1.3$), with $N_{H}\sim 4\times
10^{23}$ \cunits and a strong Fe line (equivalent width,  
 EW $\sim 800$ eV) (Smith \& Done
1996). Instead, OSSE observations (Bassani et al. 1995) showed a steeper
photon index ($\Gamma\sim 2.1 \pm 0.3$)  in the 50-200 keV energy range
suggesting the presence of a strong reflection component. \asca
observations (Turner et al. 1997, Comastri et al. 1998) showed a
flat power-law index, with a possible variation of the iron line intensity
and the absorption column density. The limited spectral bandpass
of \asca did not allow to constrain the properties of a reflection
component. Comparison between the \exosat and \asca observations show a
flux variability of a factor of about two.

\subsection{NGC7172} NGC7172 is an S0 galaxy (z=0.0087) belonging to a
compact group of galaxies (Hickson90).  It is classified as a Seyfert-2
galaxy on the basis of its optical spectrum (Sharples et al. 1984).  
\exosat obtained the first X-ray spectrum of the source yielding a photon
index of $\Gamma\sim 1.8$ (Turner \& Pounds 1989).  The same photon index
value was also found in a \ginga observation of the source (Smith \& Done
1996). The \ginga data analysis yielded a column density of $\sim 1\times
10^{23}$ \cunits, and a rather weak Fe line (EW $\sim 50$ eV).  Turner et
al. (1997) find an EW of $\sim 80$ eV. 
 The \asca data as well as combination with {\it GRO/OSSE}
 observations (Ryde et al. 1997)  show a flat spectrum of $\Gamma\sim 1.5$
 and therefore provided evidence for spectral variability 
 between the \ginga and \asca epoch. 
The 2-10 keV flux has been
fairly constant during the 1977-1989 period at a level of about $3-4\times
10^{-11}$ \funits. However, Guainazzi et al. (1998) presented evidence for
significant short term (hours) and long term (months) variability using
two \asca observations separated by a year.  They find a flux decrease by
a factor of 3 over a year. The Fe line flux decreased by a similar amount
providing important constraints for the size of the reflecting material.

\section{DATA ANALYSIS}

In this work, we present the results from the analysis of the PCA
(Proportional Counter Array) data only.  The PCA consists of five
collimated (1$^{\circ}$ FWHM) Xenon proportional counter units (PCU). The
PCU are sensitive to energies between 2 and 60 keV. However, the effective
area drops very rapidly below 3 and above 20 keV.  The energy resolution
is 18 per cent at 6.6 keV (Glasser, Odell \& Seufert 1994).  The
collecting area of each PCU is 1300 $\rm cm^2$. We use only 3 PCUs (0 to
2); the data from the other two PCU were discarded as these detectors were
turned off on some occasions. We extracted PCU light curves and spectra
from only the top Xenon layer in order to maximize the signal-to-noise
ratio. The data were selected using standard screening criteria: we
exclude data taken at an Earth elevation angle of less than 10$^{\circ}$,
pointing offset larger than 0.02$^{\circ}$, during and 30 minutes after
the satellite passage through the South Atlantic Anomaly (SAA), and when
 the particle counts {\small electron0,1,2~} are higher than 0.1.

We use the {\small PCABACKEST v2} routine of {\small FTOOLS v 4.1.1} to
generate the background models which take into account both the cosmic and
the internal background. The internal background is estimated by matching
the conditions of the observations with those in various model files. Most
of the internal background is correlated with the L7 rate, the sum of 7
rates from pairs of adjacent anode wires. However, there is a residual
background component correlated with recent passages from the SAA.
Therefore, the use of a second, activation component is also employed.
 Comparison with blank fields observations shows that 
 the above background model represents the real 
 background accurately (within 1.5 per cent in the 3-20 
 keV band). 

Due to the large field-of-view of the PCA detector, contamination of the
observed spectrum by nearby sources is likely. We have therefore checked
the \asca GIS 2-10 keV images of our sources. These show no nearby sources
(the GIS field-of-view is 40 arcmin diameter).  However, Turner et al.
(1997) point out that even the \asca spectrum of NGC4507 may be somewhat
contaminated by an adjacent point source (present in the soft \rosat PSPC
0.1-2 keV band). This source cannot be resolved within the limited \asca
GIS spatial resolution (about 3 arcmin half-power-diameter).  Possible
contamination at radii larger than the \asca GIS field-of-view can be
investigated using the \rosat PSPC images of our sources. Indeed, in
NGC7172 and ESO103-G35 we find bright sources at distances from 30 to 60
arcmin from our targets. 
 Inspection of the NASA extragalactic database (NED) 
 reveals no coincidences. 
 Therefore,  these are most probably Galactic stellar
sources and thus with soft X-ray spectra whose contamination in the \xte
band is expected to be small.

\section{THE OBSERVATIONS}

In total 43 observations with exposure time of 151.0, 46.3, 118.6, and
78.2 ksec have been obtained with \xte for ESO103-G35, IC5063, NGC4507 and
NGC7172 respectively. The period of observations ranges from about 7 days
for NGC7172, and 15 days for NGC4507, up to about 5 months for IC5063 and
7 months for ESO103-G35. The observation dates for each dataset together
with the exposure time as well as the observed background-subtracted count
rate in the full 2-60 keV PCA energy band are given in Tables 1 to 4.


\begin{table}
\begin{center}
\caption{The ESO103-G35 observations}
\label{eso103_obsn}
\begin{tabular}{ccc}
Date & Exp (ksec) & Ctr \\ \hline 
11-Apr-1997 & 14.4 &  15.38$\pm$0.11\\ 
12-Apr-1997 & 10.7 &  14.94$\pm$0.11\\
12-Apr-1997 & 15.7 & 14.68$\pm$0.10\\
13-Apr-1997 & 6.3 & 15.76$\pm$0.15\\
18-Jul-1997 & 6.2 & 12.08$\pm$0.16\\
20-Jul-1997 & 8.6 & 13.10$\pm$0.14\\
21-Jul-1997 & 11.7 & 14.12$\pm$0.11\\
22-Jul-1997 & 11.4 & 15.92$\pm$0.12\\
23-Jul-1997 & 10.4 & 15.36$\pm$0.11\\
24-Jul-1997 & 11.5 & 15.32$\pm$0.11\\
26-Jul-1997 & 12.4 & 14.30$\pm$0.11\\
13-Nov-1997 & 1.9 & 13.69$\pm$0.27\\
14-Nov-1997 & 8.7 & 12.84$\pm$0.13\\
14-Nov-1997 & 19.4 & 13.20$\pm$0.09\\
14-Nov-1997 & 1.4 & 16.10$\pm$0.32\\
\hline 
\end{tabular}
\end{center}
\end{table}

\begin{table}
\begin{center}
\caption{The IC5063 observations}
\label{ic5063_obsn}
\begin{tabular}{ccc}
Date & Exp (ksec) & Ctr \\ \hline 
22-Jul-1996 & 18.0 & 6.34$\pm$0.07 \\ 
01-Sep-1996 & 9.8 & 4.45$\pm$0.10\\
01-Sep-1996 & 1.4 & 3.42$\pm$ 0.24\\
19-Dec-1996 & 11.1 & 4.61$\pm$0.09\\
20-Dec-1996 & 6.0 & 4.73$\pm$0.12\\
\hline 
\end{tabular}
\end{center}
\end{table}

\begin{table}
\begin{center}
\caption{The NGC4507 observations}
\label{ngc4507_obsn}
\begin{tabular}{ccc}
Date & Exp (ksec) & Ctr \\ \hline 
24-Feb-1996 & 12.6 &  7.26$\pm$0.08\\ 
24-Feb-1996 & 3.3 &  6.96$\pm$0.16\\
26-Feb-1996 & 12.7 & 8.33$\pm$0.08\\
26-Feb-1996 & 14.0 & 6.99$\pm$0.07\\
27-Feb-1996 & 3.6 & 7.46$\pm$0.15\\
28-Feb-1996 & 14.5 & 6.75$\pm$0.08\\
28-Feb-1996 & 15.2 & 5.39$\pm$0.09\\
29-Feb-1996 & 0.6 & 5.83$\pm$0.35\\
29-Feb-1996 & 3.2 & 6.84$\pm$0.07\\
29-Feb-1996 & 3.2 & 6.73$\pm$0.16\\
03-Mar-1996 & 12.5 &7.70$\pm$0.08 \\
07-Mar-1996 & 16.3 & 9.58$\pm$0.07\\
07-Mar-1996 & 0.8 & 9.04$\pm$0.32\\
10-Mar-1996 & 6.1 & 7.69$\pm$0.11\\
\hline 
\end{tabular}
\end{center}
\end{table}

\begin{table}
\begin{center}
\caption{The NGC7172 observations}
\label{ngc7172_obsn}
\begin{tabular}{ccc}
Date & Exp (ksec) & Ctr \\ \hline 
18-Dec-1996 & 12.8 & 7.32$\pm$0.09 \\ 
18-Dec-1996 & 4.4 &   5.42$\pm$0.14\\
18-Dec-1996 & 5.5 &   7.61$\pm$0.12\\
19-Dec-1996 & 10.0 &  8.13$\pm$0.10\\
19-Dec-1996 & 3.2 &  8.38$\pm$0.17\\
23-Dec-1996 & 14.3 & 6.12$\pm$0.08\\
23-Dec-1996 & 9.5 & 4.84$\pm$0.09\\
25-Dec-1996 & 11.7 &2.25$\pm$ 0.09 \\ 
25-Dec-1996 & 6.6 & 2.47$\pm$ 0.11 \\
\hline 
\end{tabular}
\end{center}
\end{table}

\begin{figure*}
\includegraphics[height=10.0cm]{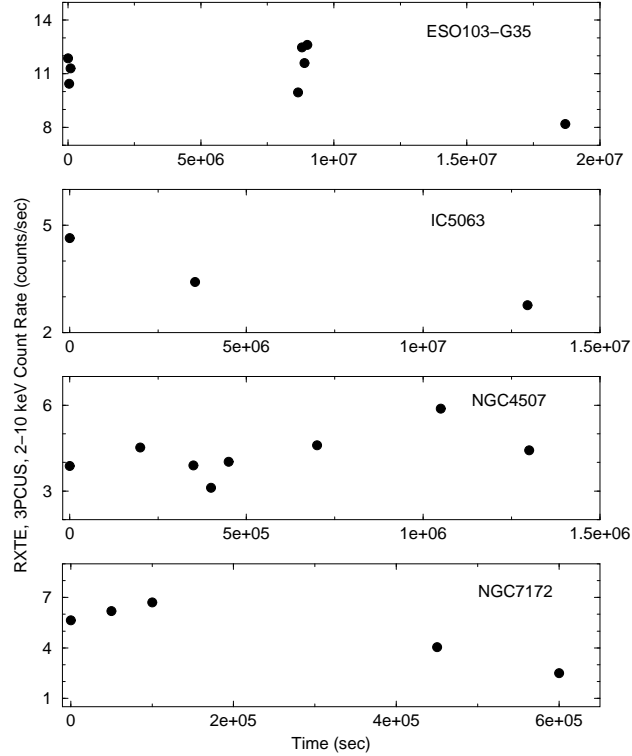}
\caption{Light curves from the PCA, 0,1,2 PCUs, 2-10 keV band for the four
sources. Bin size is 25ksec, and the time axis in seconds from the
beginning of the first observation for each source. }
\label{lcurve}
\end{figure*}

In Fig. \ref{lcurve} we 
 present the 2-10 keV light curves for all sources using a
bin size of 25ksec (time in this figure is measured from the beginning of
the first observation of each source). The light curves show variations on
all sampled timescales (days/months). For example, ESO103-G35 shows a
$\sim 30$ per cent variation within a few days, while IC5063 shows a decline by a
factor of $\sim 2$ within 4 months. The maximum amplitude of the observed
variability is $\sim 3$ in the case of NGC7172, $\sim 2$ for NGC4507 and
IC5063, and $\sim 60$ per cent in the case of ESO103-G35 light curve.
Consequently, on timescales larger than $\sim 1$ day, these four Seyfert 2
galaxies show $2-10$ keV flux variability similar to that observed in
Seyfert 1. In fact, the light curves show significant variations on time
scales as short as $\sim$ a few hundred seconds as well (these results
will be presented elsewhere). This is not an unexpected result, since for
these Seyfert 2 galaxies we believe that we can detect the nuclear
emission above a few keV directly.

\begin{figure*}
\rotatebox{270}{\includegraphics[height=10.0cm]{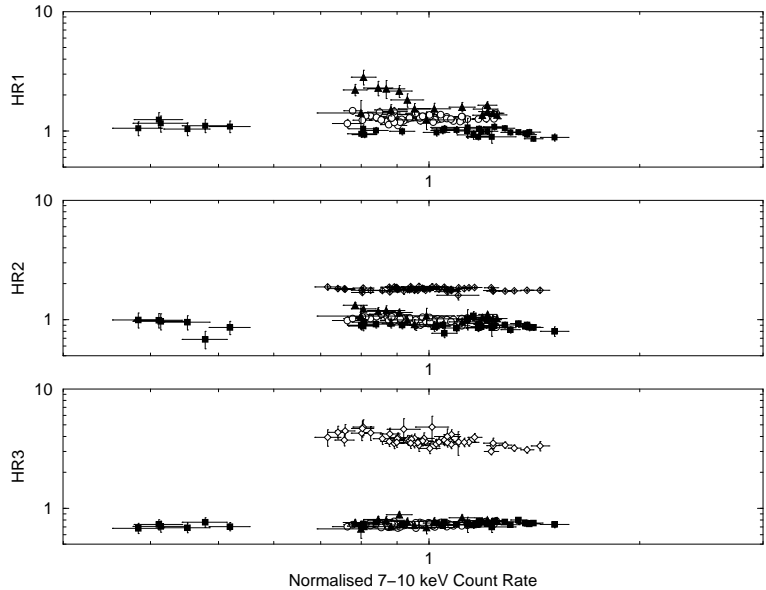}}
\caption{Hardness ratios vs normalised 7-10 keV count rate for the four
sources (for the definition of $HR1, HR2$ and $HR3$ see section 4.1). The
hardness ratios were estimated using 5400 sec binned light curves.  Open 
circles are for ESO103-G35, filled triangles for IC5063, open diamonds for
NGC4507 and filled squares for NGC7172.}
\label{ratio} 
\end{figure*}

\subsection{Hardness Ratios} 

Next, we investigate if there is any spectral variability during the flux
variations. For this reason we extracted light curves at the following
energy bands: $2-5$, $5-7$, $7-10$ and $10-20$ keV, using a bin size of
5400 sec (this is roughly equal to the orbital period of \xte). Variations
of the emission/absorption components in the spectrum of Seyfert 2
galaxies are expected to contribute in a different way in these energy
bands. For example, any absorbing column density changes around $1\times
10^{23}$ \cunits should affect mainly the $2-5$ keV band, while the Fe-K
emission line and the reflection component variations should affect the
$5-7$ keV and $10-20$ keV bands respectively. On the other hand, the
$7-10$ keV band light curve  should be representative of the
power-law continuum variations only. Using these light curves we
calculated three hardness ratios: $HR1=CR_{7-10keV}/CR_{2-5keV},
HR2=CR_{10-20keV}/CR_{7-10keV}$, and $HR3=CR_{7-10keV}/CR_{5-7keV}$, where
$CR_{E1-E2}$ is the count rate of the $E1-E2$ band light curve. Under the
hypothesis that there are no spectral changes, the $HR$s should remain
constant.

Fig. \ref{ratio} shows the $HR$ plot for the sources 
 as a function of the normalised
$CR_{7-10keV}$ (the normalisation is necessary in order to compare the
sources since they have different luminosities). NGC7172 shows the largest
amplitude variations in the $7-10$ keV band. Within the common range of
the normalised $CR_{7-10keV}$ values, the $HR$s have similar values for
all objects except for NGC4507 (open diamonds in Fig. \ref{ratio}). 
Both $HR2$ and
$HR3$ for this source have a large value. In fact we could not calculate
$HR1$ as its $2-5$ keV light curve has a mean value of almost zero. The
similarity of the $HR$ values for ESO103-G35, IC5063 and NGC7172 suggests
that they have similar spectra, while the NGC4507 spectrum appears to be
much ``harder". This is probably due to the fact that NGC4507 has a larger
absorbing column density, which reduces mainly the count rate in the
``soft" bands, and hence results in larger $HR$ values.

In order to investigate the presence of spectral variations we fitted the
data for each galaxy with a model of the form:  
$\log(HR)=a+b*log(CR_{7-10 keV})$ (in fitting the data we took account of
the errors in both $HR$ and $CR_{7-10 keV}$). In all cases, a line in the
log-log space fits the data well (ie the best fit \x2 values can be
accepted with confidence larger than $5$ per cent).  If the $HR$ values remain
constant during the flux variations, we expect $b$ (ie the slope of the
line) to be consistent with zero. We discuss below the best fit results
for each galaxy.

ESO103-G5 shows no $HR1$ variations ($b_{HR1}=0.05\pm0.06$). On the other
hand, both $HR2$ and $HR3$ are variable: $b_{HR2}=-0.23\pm0.06$ and
$b_{HR3}=-0.10\pm0.04$. The $HR2$ variations suggest that the spectrum of
the source above $7$ keV steepens as the source brightens.  The $HR3$
ratio shows the opposite behaviour. It implies a ``flattening" of the
$5-10$ keV spectrum as the source brightens. In the case of IC5063, both
$HR1$ and $HR2$ are variable ($b_{HR1}=-1.0\pm0.2$, and
$b_{HR2}=-0.40\pm0.10$) while $HR3$ remains constant ($b_{HR3}=0.06\pm
0.07$).  Both $HR1$ and $HR2$ indicate that the source's spectrum becomes
softer as it brightens. NGC4507 shows a different behaviour. $HR2$ remains
constant ($b_{HR2}=-0.04\pm0.03$), while $HR3$ is significantly variable
($b_{HR3}=0.45\pm 0.08$). The spectrum below $10$ keV appears to become
softer as the source flux increases. NGC7172 shows a similar behaviour.  
$HR2$ remains constant ($b_{HR2}=-0.04\pm 0.03$), $HR3$ shows marginal
changes ($b_{HR3}=0.06\pm0.03$) while $HR1$ shows more significant
variations ($b_{HR1}=-0.12\pm0.4$). These variations indicate again a
steepening of the spectrum below $10$ keV as the source flux increases.

\section{SPECTRAL ANALYSIS}

In our spectral analysis, we used only data between 3 and 20 keV, where
the effective area of PCA, and hence the signal-to-noise ratio, is
highest. By discarding the data below 3 keV we also avoid the
complications associated with the soft X-ray excess in these sources (see
Turner et al. 1997). The PCA data from each observation were binned to
give a minimum of 20 counts per channel (source plus background).  The
spectral fitting analysis was carried out using the {\small XSPEC v.10}
software package on the basis of ``joint fits'' to our {\sl RXTE}
observations. All errors correspond to the 90 per cent confidence level
for one interesting parameter. All energies quoted refer to the emitter's
rest-frame.

We have employed two group of models (see below) in order to characterize
the spectra of the objects at each observation. Within each group, in
order to assess the significance of new parameters added to the initial
model of the group we have adopted the 95 per cent confidence level using
the F-test for additional terms. To compare models with different number
of parameters within each group and models between the two groups we have
used the ``likelihood ratio" (Mushotzky 1982) and accepted values of the
ratio larger than 100 as showing significant improvement in the model
fits.

Note that the \xte background estimation software calculates errors on the
simulated background spectrum assuming Poisson statistics. This is an
overestimate of the statistical error on the background spectrum, which is
estimated using a large amount of data. In fact, for faint objects (such as
AGN) the error that the standard software calculates for the background is
comparable to the error in the total observed rate. In practice, the
background spectrum errors should be negligible compared to the errors on
the observed spectrum. Since the {\small XSPEC} software estimates the
error in the net spectrum (source - background) by combinning the errors
in the observed spectrum with those of the simulated background, the
reduced \x2 values in many cases are very small. However, the F and
``likelihood ratio'' tests are not affected by this. Finally, due to the
large errors, the confidence regions that we estimate for the best fit
parameter values are almost certainly enlarged.

\subsection{Power Law and Gaussian Line Models.} 

Following previous \ginga and \asca results, we first employ a simple
spectral model consisting of a power-law continuum, with photon spectral
index $\Gamma$, modified by absorption in a column density, $N_H$, of cool
neutral material. A Gaussian line was also included to account for Fe-K
emission.  The line energy and the width were allowed to vary freely.  
However, when the width was found to be smaller than the \xte spectral
resolution, we fixed it at $\sigma_{line}=0.1$ keV.  The values of the
$N_H$, the photon index and the normalization of the Fe line are free
parameters but tied to the same value for all observations. The
normalization of the power law is allowed to vary freely.  The results of
fitting this spectral model are presented as model A in Tables 5, 6, 7 and
8.  Next, we allow for the column density and the power-law index to vary
freely (models B and C in the same tables). Finally, in model D we allow
for the normalization of the Gaussian line to vary freely while now both
the power-law index and the column density are tied to the same value
between all the observations.  We discuss the results for each galaxy in
turn.

{\bf ESO103-G35.} 
It shows a large column density ($N_{H}\sim 1.4\times
10^{23}$ \cunits) in excellent agreement with the \ginga observations.  
However, the spectral index is flat ($\Gamma\sim 1.6$, model A) in
contrast with the \ginga results. Moreover, the best fit energy line is
very low ($E \sim$ 6 keV) and marginally resolved, ie broader 
 than the energy resolution of PCA, having a $\sigma$ of $\sim
0.7$ keV. The EW of the line does not remain constant. It ranges from
$490^{+41}_{-38}$ to $661^{+56}_{-51}$ eV.  The data show strong evidence
for spectral variability, since models B,C and D give a significantly
better fit to the data when compared to model A. When the line
normalisation is untied (model D), we obtain a $\Delta \chi^{2}=68$ for 14
addtitional parameters. A larger $\Delta \chi^{2}$ value is obtained when
the column density is untied and an even larger $\Delta \chi^{2}$ when the
power law index is untied ($\Delta \chi^{2}=107$ and 162 respectively for
14 additional parameters). Based on the likelihood ratios, model C gives
the best fit to the data.

Following Forster
et al. (1997), we tried an additional model. We added an absorption edge
component (model {\small EDGE} in XSPEC) to model A and kept the Fe line
energy fixed at 6.4 keV.  We obtain a highly statistically significant
improvement in \x2 compared to model A ($\Delta\chi^2\sim 143$ for one
additional parameter).  The photon index becomes now steeper ($\Gamma=
1.68^{+0.03}_{-0.02}$ and $N_H= 1.7^{+0.5}_{-0.5}\times 10^{23}$ \cunits). The
best-fit energy for the edge is $6.8^{+0.10}_{-0.10}$ keV and the
optical depth is $\tau= 0.2 \pm 0.02$.

{\bf IC5063.} Model A does not provide a good fit in this case.  The
column density is $2\times 10^{23}$ \cunits while the power-law is flat
($\Gamma \sim 1.6$) both in good agreement with the \ginga results. The EW
of the Fe line varies between $298^{+26}_{-45}$ and $462^{+55}_{-70}$ eV.
As before with ESO 103-G35, we find highly significant evidence for
spectral variability.  Models B and C give a significant \x2 improvement
compared to model A (in this case model D does not improve the fit
significantly). The best fit is obtained when the photon index is untied
(model C):  $\Gamma = 1.15 - 1.70$. However, the Fe line energy is low
($E\sim 6.12$ keV with $\sigma \sim 0.5$). We therefore included an
absorption edge in model C.  The line energy is marginally improved,
$E=6.19^{+0.10}_{-0.12}$ keV, with $\sigma=0.48$.  The edge energy was
fixed at 6.9 keV while the optical depth found was $\tau=0.07$. However,
the inclusion of the edge did not improve the fit at a statistically
significant level ($\chi^2=175.2/235$, as compared to $\chi^2=177.2/236$ 
for model C).

{\bf NGC4507.} Model A results in a high absorbing column ($\sim 4.5\times
10^{23}$ \cunits) and a flat spectrum ($\Gamma \sim 1.4$), in agreement
with the \ginga observations.  The line energy is $\sim 6.16$ keV with the
line width being unresolved ($\sigma \sim 0.33$). The EW ranges from
$353^{+47}_{-34}$ up to $655^{+89}_{-62}$ eV. All models B,C and D give a
significant \x2 improvement when compared to model A, indicating again the
presence of significant spectral variability. Based on the likelihood
ratios, it is model B this time that provides the best fit to the data.

{\bf NGC7172.} Model A shows a large column density ($N_{H} \sim 1.1\times
10^{23}$ \cunits) in agreement with previous \ginga results. The power-law
has a value of $\Gamma\sim 1.8$, also in good agreement with the \ginga
results, and typical of the intrinsic AGN spectral index.  The Fe line has
an energy of $\sim 6.24$ keV. The line width is consistent with the line
being narrow.  The resulting  EW varies from $96^{+13}_{-22}$ to
$360^{+49}_{-28}$ eV in the nine observations.  Model D does not
improve the fit, while models B and C give a statistically significant
improvement to the \x2 value of the fit.  When the column density becomes
a free parameter (model B) we obtain a $\Delta \chi^2 \sim 30$ for 12
additional parameters. When the photon index becomes untied (model C), we
obtain an equally good \x2 with $\Gamma$ varying between $1.7$ and $2.0$.
This suggests that the two parameters (photon index and column density)
are degenerate and it is therefore difficult to disentangle the origin of
the observed spectral variation. We plot the variations of column density
and of the photon index as a function of the $3-20$ keV flux in Fig.
\ref{nh} and \ref{gamma} respectively. We can see that the changes in the
spectrum of this source can be explained by a decrease of $N_{H}$ by 
 $\sim 30$ per cent as the source flux increases by a factor of
$\sim 3.5$. Alternatively, a $\Gamma$ increase by $\sim 0.3$ (from $\sim
1.65$ up to $\sim 1.95$) can explain the spectral changes equally well.

\subsection{Power Law, Gaussian Line and Reflection Component Models.}    

Although the power-law plus gaussian line prescription defined above gives
an acceptable fit in terms of the $\chi^{2}$ statistic, there is evidence
from the flat power-laws derived for NGC4507, IC5063 and ESO103-G35 for
the presence of a strong Compton reflection component. The next step in
the current analysis was therefore to include a reflection component in
the spectral modeling. Specifically we use the {\small PEXRAV} model
(Magdziarz \& Zdziarski 1995)  in {\small XSPEC}.  This model calculates
the expected X-ray spectrum when a source of X-rays is incident on
optically thick, neutral (except hydrogen and helium) material. We assume
that both the reflection component and the power-law are absorbed by the
cold gas column density. The strength of the reflection component is
governed by the parameter R, representing the strength of the reflected
signal relative to the level of the incident power-law continuum. We set
R=1 (which corresponds to a $2\pi$ solid angle subtended by the optically
thick material). We fix the inclination angle for the disk at
$i=60^{\circ}$, since the shape of the reflection spectrum below 20 keV is
relatively independent of the inclination angle. We also fix the energy of
the exponential cutoff at 300 keV. Both the iron and light element
abundances were kept fixed at the solar abundance values.

We added this reflection component to all models A,B,C and D, creating a
new set of models (A',B',C' and D').  In all new models, the normalisation
of the reflection component is tied to a single value across the set of
observations. We also set the spectral index of the power-law continuum to
$\Gamma=1.9$ (in all models apart from model C').  As before, in model A'
the power-law normalisation is allowed to vary in each observation while
 the column density and the Fe line normalization are
tied between the different observations. In model B' we untie the column
density, in model C' we untie the photon index and in model D' we let the
gaussian line normalisation free. Finally, we let the reflection component
normalisation free (model E'). The results are presented in tables 9 to
12.

{\bf ESO103-G35.} In the case of ESO103-G35, the inclusion of a reflection
component gives a statistically significant improvement to the data fit.
For example, we obtain a $\Delta \chi^2 = 163$ (with the addition of one
extra parameter) when we include the reflection component in model A.  
This is in agreement with the findings of Smith \& Done (1996)  using
\ginga data. Furthermore, all models (B',C',D' and E') significantly
improve the fit when compared to model A'. The best fit to the data is
obtained when the photon-index is allowed to vary with time (model C',
$\Gamma\sim 1.8-2$). An equally good fit is also obtained when the
reflection component normalisation is allowed to vary with time (model
E'). Both these models give a better fit than model C, showing again that
the inclusion of the reflection component significantly improves the fit.

\begin{figure}
\includegraphics[height=7.0cm]{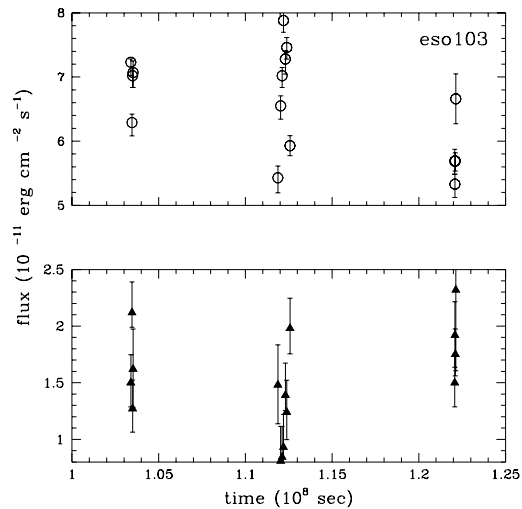}
\caption{ESO103-G35: 
The 3-20 keV flux of the power-law (upper panel) and of the reflection
component (lower panel)  as a function of  time from the mission's launch
(90 per cent  errors are plotted)}
\label{eso103_pexrav}
\end{figure}

\begin{figure}
\includegraphics[height=7.0cm]{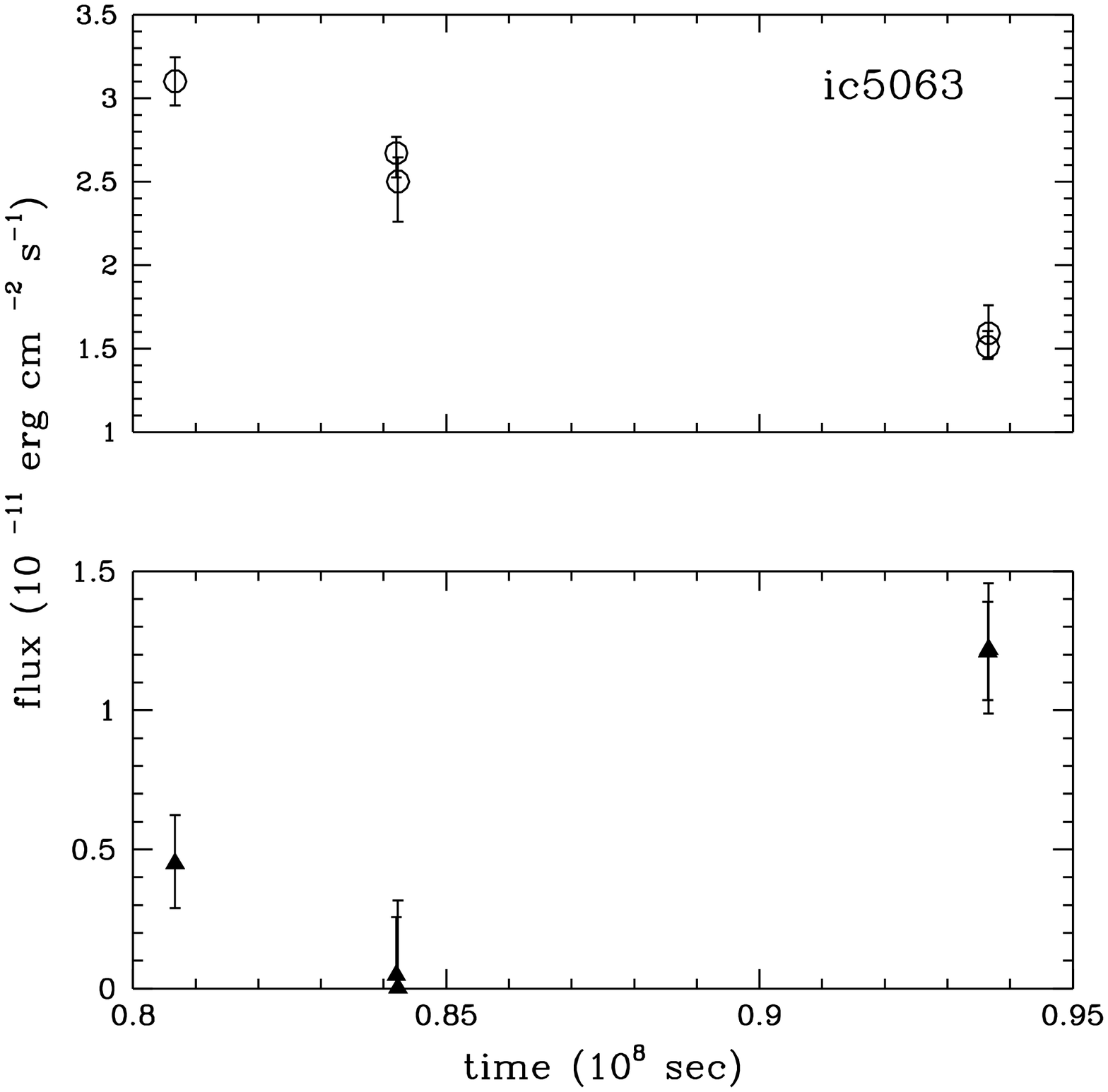}
\caption{IC5063: the 3-20 keV flux of the power-law (upper panel) and
of the reflection component (lower panel)  as a function of  time from the
mission's launch (90 per cent errors are plotted).}
\label{ic5063_pexrav}
\end{figure}

In Fig. \ref{eso103_pexrav} we present 
the variations of the power-law and the reflection
flux between the different observations (model E'; time in this figure and
in Fig. \ref{ic5063_pexrav} as well is measured in seconds from 
the mission launch date). We
see that the power-law component shows rapid flux variations (up to a
factor of 50 per cent) within days. Similar variations, but not well
correlated with the power-law variations, are observed in the reflection
component flux as well. In Fig. \ref{gamma} 
we plot model's C' best fit $\Gamma$
values as a function of the $3-20$ keV flux. A change of $\sim 0.15$ (from
$\Gamma \sim 1.8$ up to $\Gamma \sim 2$) can explain the spectral
variability in this case.

However, the best fit line energy in models C' and E' is abnormally low
($E\sim 6.15$ keV). We therefore introduced an absorption edge to model
E'. The resulting \x2 is 505/714, which indicates that, compared to model
E', the improvement to the fit is statistically significant. The best fit
line energy now becomes $6.23\pm 0.10$ keV, roughly consistent with the
line energy of neutral Fe. The best fit edge energy is $E_{edge}=7.69
 \pm 0.14$ keV (ionised Fe) while the optical depth is $\tau=0.1\pm 0.01$.
Similar results are obtained when we add a warm edge component to model
C'.

{\bf IC5063.} In IC5063 the inclusion of a reflection component does not
represent a statistically significant improvement. Models C', E' and C
(which does not include a reflection component) give an equally good fit
to the data.  However, the inclusion of the reflection component (for
example model E')  yields a line with an energy of $E=6.25\pm0.10$
keV, which is roughly consistent with the energy of line from neutral Fe.
In Fig. \ref{ic5063_pexrav} the variations of the power-law 
and the reflection flux are
presented (model E').  The power-law component flux decreases by a factor
of two with time. In contrast, the reflection signal, although variable as
well, does not correlate with the power-law flux.  Alternatively, a change
of $\Gamma$ between roughly $1.6$ and $2.2$ (model C', see Fig. \ref{gamma})
yields a similarly good fit to the data, and results in the same best fit
line energy value. However, the photon index does not correlate well with
the continuum flux.
  
{\bf NGC4507.} In this case, the use of a reflection component improves the
fit dramatically. Models B' and C' give a much better fit to the data than
model B. These two models (untied column densities or photon indices) fit
equally well the data: in the first case the $N_H$ varies roughly between
$4.8$ and $5.3\times 10^{23}$ \cunits, while in the second case $\Gamma$
varies between $1.9$ and $2.0$.
   
\begin{figure*}
\includegraphics[height=8.0cm]{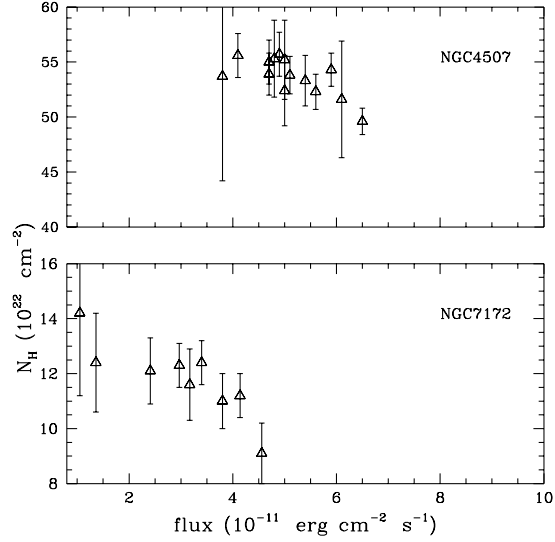}
\caption{The variation of the column density in NGC4507 (model B' with edge) 
and NGC7172 (model B) as a function of the 3-20 keV flux}
\label{nh}
\end{figure*}

\begin{figure*}
\includegraphics[height=8.0cm]{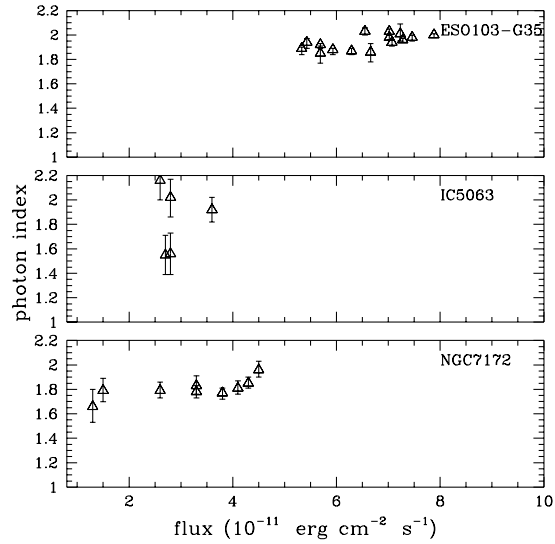}
\caption{The variation of the photon index in ESO103-G35 (model C' with edge),
 IC5063 (model C') and NGC7172 (model C) 
 as a function of the 3-20 keV flux}
\label{gamma}
\end{figure*}

The line energy is again low ($E\sim 6.2$ keV) and therefore we included
an absorption edge component in both models B' and C' above. We obtain a
statistically significant improvement in the fit in both cases. However,
the best fit is now clearly obtained in the case of a column density
variation (\x2=643/794 as compared to \x2=655/794 for an untied $\Gamma$).
The edge energy was frozen at 6.9 keV while its optical depth was found to
be $0.16\pm0.02$. The line energy was kept fixed at 6.4 keV ($\sigma$=0.1
keV). In Fig. \ref{nh} we plot the variations of $N_{H}$ as a function of
the 3-20 keV flux (model B' with edge). A $\sim 10$ per cent decrease in
the absorpting column while  the source flux increases by a factor of 2
can explain the observed spectral changes.

{\bf NGC7172.} In this case, the inclusion of a reflection component does
not improve the fit at a statistically significant level. Models B' and C'
present the best fit to the data, however the goodness of fit is
comparable to that of the models B and C. The lack of a reflection
component is rather expected given the steep power-law index found earlier
in the case of the simple power-law and a Gaussian line model. Note that
the \ginga results required the presence of a reflection component only
marginally.

\subsection{Soft excess models} 
The X-ray spectra of Seyfert-2 galaxies
usually show some level of soft X-ray emission despite the strong
photoelectric absorption (eg Turner et al. 1997). This soft excess could
be due to either a strong star-forming component or scattered emission
from the nucleus. The contribution of the soft excess in our case (ie
above 3 keV) should be negligible. Nevertheless, we tried to investigate
whether a constant soft excess combined with a power-law component with
varying normalization may explain the strong spectral variability observed
in our data. Therefore, we include a power-law component with the same
photon index as the hard X-ray power-law and with a free normalization
(but tied between all the observations) in the models above.

In NGC7172 we add the soft power-law to model A since a reflection
component is not needed by the data. We obtain \x2=303.8 ie identical to
that of model A, despite the addition of one extra parameter. We conclude
that an extra soft component is not needed by the data. In the other three
cases we add the soft power-law component to model A'. In the case of
ESO103-G35 we obtain \x2=568.5/729 which represents a statistically
significant improvement to model A'. However, the fit is worse when
compared to models C' and E'. In NGC4507 we obtain $\chi^2=738/806$
similar to model A' with the soft power-law normalization being zero.  
Finally, in the case of IC5063 we obtain an identical $\chi^2$ to model A'
(269/239) while the soft power-law normalization is again zero. The above
confirm that the presence of any soft excesses do not affect our data
above 3 keV, and they cannot account for the observed spectral
variability.
 
\section{SUMMARY}

{\bf ESO103-G35} presents a complex spectrum. Models C'and E', with the
addition of a warm edge, give the best fit to our data. Our best fit edge
energy is $7.69\pm0.14$ keV with $\tau=0.1\pm 0.01$. The best fit edge energy
matches the results from the 1988 and 1991 \ginga observation (Warwick et
al 1993, Smith and Done 1996), and also the 1996 \asca observation
(Forster et al 1999). The energy of the edge implies the presence a
significant amount of ionised iron. The optical depth of the iron edge was
observed to increase between the 1994 and 1996 \asca observations (Forster
et al 1999), and the edge line in the former observation was consistent
with that from neutral iron. Our best fit edge optical depth is
significantly smaller than the respective \asca result (Forster et al.
1999).  However, the quality of the present data set is not good enough to
investigate if there exist any changes in the optical depth and/or the
energy of the edge during the \xte observations. The spectral variability
can be explained by a variable spectral index (model C', Fig.
\ref{gamma}).  Although these variations are subtle, they are highly
statistically significant. The photon index is correlated with the 3-20
keV flux in the sense that the spectrum steepens with increasing flux. The
same variations can be equally well explained by a variable reflection
component (model E', Fig. \ref{eso103_pexrav}).  
Interestingly, Akylas et al. (2000) find no
reflection component in the \sax observations despite the large passband
($\sim$2-60 keV) and the good signal-to-noise 
 ratio of their observations. 
This may suggest that the relative normalization of the
power-law and the reflection component must change in different epochs,
and supports the interpratation that the spectral variations seen by \xte
are caused by a variable reflection component. Furthermore, in this case
we can also explain naturally the fact that although the $HR2$ ratio of
ESO103-G35 is variable, the $HR1$ ratio remains constant.

{\bf IC5063} can be well fit with a simple spectrum consisting of a
power-law and a Gaussian line but the energy of the line is abnormally low
($\sim6.12$ keV).  The addition of an edge does not yield a reasonable value
for the energy of the Fe line.  However, the addition of a reflection
component (models C' and E'), although it does not improve the \x2, yields
an Fe line with an energy of $6.25\pm0.1$ keV compatible within the errors
with the energy of cold Fe line. IC5063 shows strong evidence for spectral
variability (Fig. \ref{ratio}).  The spectral variability can be explained
either by the reflection component being variable between the observations
(model E') or by the photon index being variable (model C', $\Gamma \sim$
1.6 - 2.2). According to model E' best fit results there is no correlation
between the reflection component and the intrinsic power law flux
variability (Fig. \ref{ic5063_pexrav}).  
Similarly, in the model C' case, there is no 
 clear correlation between the photon index and the continuum flux
(Fig. \ref{gamma}).

{\bf NGC4507} presents a similar spectrum to ESO103-G35. We find strong
evidence for a reflection component (note that such a component was not
evident in neither the \ginga nor the \asca data). We also find evidence
for a cold Fe edge (with an energy fixed at 6.9 keV), in excess of the
absorption edge associated with the reflection component. Assuming the
iron cross section given by Leahy and Creighton (1993), the optical depth
of the edge ($\tau=0.16\pm 0.02$) implies an equivalent hydrogen column
density of $\sim 2\times 10^{23}$ \cunits. This is smaller than our best
fit $N_{H}$ value of $5\pm1.0 \times 10^{23}$. The edge component is $not$
required by the data, but with its addition we can get a good fit to the
data with the iron line energy fixed at 6.4 keV.  However, there is a
possibility that residuals in the PCA calibration matrix around 5 keV due
to Xe may still be present. In such a case it is possible that the line
energy obtains somewhat low values due to these residuals.  Therefore, we
believe that the detection of a cold Fe edge in this object remains
uncertain. We find strong evidence for spectral variability (Fig.
\ref{ratio}).  The best model fit to the data (model B') suggests flux
correlated column density changes: the column becomes lower with
increasing flux (Fig. \ref{nh}).  This model can explain naturally the
$HR3$ variations observed for this object, and at the same time the lack
of $HR2$ variations.

{\bf NGC7172} is the only object which shows no evidence for a reflection
component. \ginga found marginal evidence for such a component while in
\sax data a reflection component is needed only if the spectral index is
fixed to the canonical $\Gamma=1.9$ value (Akylas et al. 2000). Similarly,
if $\Gamma$ is fixed to 1.9, models with reflection (B' and C') can fit
the present data set well but simple power law models (B and C) can fit
the data equally well. There is no evidence for an edge, and spectral
variability is observed again.  The spectral fitting cannot discriminate
between variations in the photon index or the column density: as the flux
increases, either the column drops ($14 - 11\times 10^{23}$\cunits) or the
photon index steepens ($1.66 -1.96$) in a similar fashion to ESO103-G35
and NGC4507.

\section{DISCUSSION AND CONCLUSIONS} 

The \xte observations of ESO103-G35, IC5063, NGC4507 and NGC7172 show
that, in all objects, the X-ray flux is significantly variable on time
scales larger than $\sim$ a day, with an amplitude similar to that
observed in Seyfert 1 galaxies. This result on its own suggests that, in
these objects, we are seeing directly the nuclear emission. This emission
is transmitted through a heavily absorbing material with a density of
$\sim 1-5 \times 10^{23}$ \cunits. Apart from the flux variability, we
searched for the presence of spectral variations, taking advantage of the
fact that the present data set consists of many observations for each
object all made with the same instrument. In the
past, spectral variability studies in Seyfert 2 galaxies were not
conclusive mainly because it is difficult to compare the results from
different spectral fits performed across various instruments with
different bandpasses. We detected significant, flux related variations in
the hardness ratios that we calculated for all objects (Fig. \ref{ratio}). 
This result suggests that spectral variability may be a ubiquitous feature in
Seyfert-2 galaxies.

In principle, the observed hardness ratio variations could be an artefact
of poor background subtraction. However, the $HR$ variations do {\it not}
show the same behaviour in all sources, and are well correlated with the
source flux. Therefore, if these variations are due to background
subtraction errors, those errors should be different at different epochs
and should operate in such a way so that the resulting ratios should mimic
the flux correlated variations seen in Fig. \ref{ratio}. We consider this
possibility unlikely to be the cause of the spectral variations that we
observe.

Still, the explanation for the observed spectral variability is model
dependent and is based on the spectral fitting results. In some cases,
these results are somehow uncertain. For example, in NGC7172 and NGC4507,
significantly different $N_{H}$ values are observed in just one
observation, namely the one with the highest flux (Fig. \ref{nh}).
Although the hardness ratios clearly show that the variability in both
NGC4507 and NGC7172 is attributed to more than one observation, we
investigated whether the spectral fitting results could be affected by
problems associated with the background subtraction in the highest flux
observations. Hence, we examined the data below 3 keV in layer 1 where the
signal is mostly due to particles as the effective area of the PCA is
negligible. We also examined the signal in the full energy band of layers
2 and 3 which are again sensitive to particles only.  We find that the
particle spectrum was consistent with the background model predictions,
especially below 5 keV. In conclusion, we believe that the spectral
variability is real in all four objects, and that our model fit results
can be used in order to explain its origin.

The origin of the spectral variability appears to be different in each
object. We discuss below briefly the implications of our results on the
processes that operate in the nuclei of Seyfert 2 galaxies.

\subsection{Photon index changes} 

In three cases, namely ESO103-G35, IC5063 and NGC7172, models with a
variable photon index (model C' for the first two sources and model C for
NGC7172) can fit the data well. Turner et al (1998) have found significant
photon index variability in many Seyfert 2 galaxies by comparing
historical X-ray observations. Significant photon index variability has
also been observed in many Type 1 objects (eg MCG 6-30-15, Lee et al.
1999; NGC4051, Guainazzi et al. 1996; Mrk 766, Leighly et al. 1996;
NGC4151, Warwick et al. 1996). In all these objects, the power law becomes
steeper with increasing flux. The same pattern is followed by the photon
index with flux in our case as well. In ESO103-G35 and NGC7172 we observe
$\Gamma$ increasing by $\sim 0.2$ and $\sim 0.3$ for a $\sim 1.6$ and
$\sim3$ flux increase respectively. In IC5063, $\Gamma$ increased by $\sim
0.6$ while the source doubled its flux, but is far from clear in this case
that the $\Gamma$ variations are correlated with the source flux.

According to standard Comptonization models, hard X-rays in AGN are
produced through the scattering of the soft UV photons from the accretion
disk on a hot ($>$ 40 keV) corona above the accretion disk. Opacity
changes in the corona can give rise to significant spectral variability,
even if the total luminosity of the hard X-rays remains constant (Haardt
et al 1997).  If the corona opacity is not dominated by pairs, as the
optical depth of the corona, $\tau$, increases the intrinsic spectrum
becomes steeper (as long as $\tau < 1$). As a result, as long as $\Gamma <
2$, the the photon index correlates with the photon flux: a $\sim 2.5$
increase in the $2-10$ keV flux can result in a $\Delta\Gamma\sim 0.4$
increase in the photon index (Fig. 7 in Haardt et al 1997). The best fit
$\Gamma$ values that we find are less than $2$ in almost all cases and
the $\Delta \Gamma$ changes that we detect for the observed source flux
variations are in broad agreement with the predictions of this model.

We note in passing that Seyfert 2 galaxies with ``flat spectrum" (ie Smith
\& Done 1996) may not be intrinsically flat. If those sources also exhibit
flux related photon index variations, then values as flat as $\sim 1.5 -
1.6$ can be expected according to Haardt et al. (1997). Monitoring
observations of these sources could resolve the issue.

\subsection{Absorption column changes} In NGC4507 and NGC7172, models with
a variable absorption column density (models B' and B respectively) give a
good fit to the data. In fact, model B' gives the best fit to the NGC4507
data. In both cases, the column density is anticorrelated with the X-ray
flux: a $\sim 5\times 10^{22}$ \cunits decrease in $N_{H}$ is observed
while the source flux increases by a factor of $\sim 2$ and $\sim 3$
respectively (Fig. \ref{nh}). Absorbing column variations in Seyfert 2 galaxies
have been observed in the past as well (eg NGC526A, Turner \& Pounds 1989;
NGC7582, Warwick et al. 1993; ESO103-G35, Warwick, Pounds and Turner 1988).
In the first two cases significant column changes on a time scale of years
were detected, while two \exosat observations of ESO103-G35 revealed a
significant drop in the column density within 90 days.

The flux correlated $N_{H}$ variations can be explained if the obscuration
is not due to a uniform torus, but due to individual clouds rotating
around the nucleus instead. We may expect clouds of different column
density to intercept the emission from the nucleus at different epochs. In
this case, a small increase of the column density will be associated with
a decrease in the total flux. This model requires both high velocities for
the covering clouds and a small X-ray source size. For example, in the
case of NGC7172, the lowest flux observations (last two points in Fig. 
\ref{lcurve})
have similar $N_{H}$ values and are $\sim 4.5$ and $\sim 5.5$ days apart
from the highest flux observation (third point in Fig. \ref{lcurve}). 
Assuming that a
cloud of $N_{H} \sim 3-5\times 10^{22}$ \cunits and velocity of $\sim
10^{4}$ km/sec covered the X-ray source in $4.5-5.5$ days, then the size
of the X-ray source should be smaller than a few ($\sim 3.5-4.5$)
light hours.

The $N_{H}$ variations could also be explained by partial photo-ionisation
of the obscuring screen. Varying degrees of photo-ionisation caused by
intrinsic luminosity variations of the source can result in column density
changes. This model can explain succesfully the variable X-ray absorption
in MR2251-178 (Halpern 1984, Pan et al. 1990). Recently, Comastri et al.
(1998), presented strong evidence for the existence of photonionised
material with a column density of $\sim 8\times 10^{22}$ \cunits in
NGC4507. In order for the absorbing column to adjust itself instantaneous
to the continuum flux variations (and hence explain the rather
``continuous" variations shown by its $HR2$ plot in Fig. \ref{ratio}), 
the absorbing
material should be close to the central source and in a thin shell (so
that its density will be larger than $\sim$ a few $10^{8}$ cm$^{-3}$,
Netzer 1993). Interestingly, Turner et al. (1998) have recently presented
evidence for the existence of absorbing material close to the central
source in Seyfert-2 galaxies. Neither the \asca nor the \xte data show
evidence for a warm edge in the spectrum of NGC4507. For an ionised
material with a column density of $\sim$ a few $\times 10^{23}$ \cunits the
respective edge should have an optical depth of less than 0.1, and hence
difficult to detect. On the other hand, ESO103-G35 shows a warm iron edge
in its spectrum but no indication of absorption column variations. Perhaps
in this case, the small amplitude continuum variability together with
possible variability of other spectral components (ie $\Gamma$, reflection
component normalisation) render the detection of any absorption variations
difficult. Finally, in the case of NGC7172, ASCA observations did not show
any evidence of photo-ionised material (Guainazzi et al 1998). The $2-10$
keV flux of the source during the ASCA observation was $\sim 1.3\times
10^{-11}$ erg cm$^{-2}$s$^{-1}$, which is similar to the flux during the
last \xte observation.  Their best fit $\Gamma$ value $1.5\pm0.15$, which
is consistent with our estimate ($1.66\pm 0.15$, Table~8).  This result
supports the idea that in NGC7172 flux related $\Gamma$ changes (ie model
C) rather than $N_{H}$ variations (model B) are responsible for the
spectral variability.

The detection of $N_{H}$ variations in just one source indicates that
distribution of the absorbing material in Seyfert 2 galaxies may not be
the same in all objects (eg Turner et al. 1998). Material away from the
nucleus is not expected to be ionised and will not respond to the
continuum variability. As a result, $N_{H}$ in these objects will remain
constant with time.

\subsection{Reflection component and Iron Line}

Model E' (a variable reflection component) can explain
the spectral variability observed in ESO103-G35 and IC5063. If true, this
would imply that the cold material responsible for reflection is close to
the nucleus eg an accretion disk as suggested by Nandra et al. (1997) for
Seyfert 1 galaxies. In this case we expect a positive correlation and no
appreciable lags in the reflection response to the continuum flux
variations. Contrary to this, we find that the reflection variations are
not correlated with the continuum variations, especially in the case of
IC5063 (Fig. \ref{eso103_pexrav}, \ref{ic5063_pexrav}). 
 One can assume, rather arbitrarily, that there is after
all a lag and therefore the reflection component is responding to past,
unobserved continuum variations. Further evidence for the existence and
possible variability of the reflection component can be provided by the
iron line results.

\begin{figure}
\rotatebox{0}{\includegraphics[height=7.0cm]{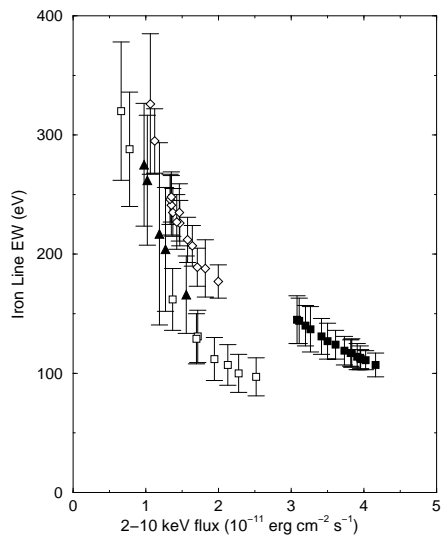}}
\caption{
The equivalent width of the 6.4 keV line plotted 
 for all objects as a function of the 2-10 keV flux. Solid squares 
 are for ESO103-G35, solid triangles for IC5063, open diamonds for NGC4507
 and open squares for NGC7172 (90 per cent errors are plotted)} 
\label{ew}
\end{figure}

All sources show a strong, unresolved emission line at $\sim 6.4$ keV. The
only exception is ESO103-G35 where the line may be marginally 
  resolved($\sigma \sim 0.5$ keV, models C' and E').
  This is consistent with the results of Turner et
al. (1997) for this source. The intensity of the line though remains
constant. This is indicated by the fact that model D' (a variable iron
line) does not improve the fit when compared with model A' in IC5063 and
NGC4507. The same holds for model D when compared with model A for
NGC7172. Fig. \ref{ew}, shows a plot of the line EW as a function of the
source
$2-10$ keV flux. The equivalent widths were estimated using results from
the best fit of model B' for NGC4507 and NGC7172, model C' for IC5063 and
the same model plus a warm iron edge for ESO103-G35. Since all models
require a constant line, in all cases the EW decreases as the
source flux increases. The errors on the EW are small enough to see that
the EW variations are significant.

The fact that the lines are narrow (except for ESO103-G35) and
constant implies that they are produced by the obscuring torus. For
IC5063, NGC4507 and NGC7172, the EW of the iron line can be as large as
$\sim 300$ eV in the lowest flux observations (Fig. \ref{ew}). The
measured column densities in the line of sight toward those sources is
$\sim 3\times 10^{23}$, $\sim 5\times 10^{23}$ and $\sim 1\times 10^{23}$
\cunits respectively. For these $N_{H}$ values, we expect an emission line
with equivalent width of $\sim 120$ eV, $\sim 200$ eV and $\sim 40$ eV
respectively (Ghisellini et al. 1994). The larger than expected EW
suggests that there exists a source of line photons other than from the
line-of-sight absorbing material.  A 6.4 keV iron line can also be
produced by fluorescence from an acrretion disk. The expected
angle-averaged EW in this case is $\sim 110$ eV. Consequently, in NGC4507
and IC5063 the iron line that we observed is probably the sum of two lines
- most of the line is produced by the torus and there is additional
contribution by the accretion disk itself. Any variations of the accretion
disk line would be diluted by the presence of the torus line and hence
difficult to detect. Furthermore, the lack of line variability indicates
that the torus should be in a distance larger than approximately 
 two light weeks and approximately 
  six light months in NGC4507 and IC5063 respectively.

For NGC7172, the discrepancy between the EW observed in the two lowest
flux observations of this source ($\sim 300$ eV), and the EW expected in
the case of a torus plus an accretion disk line ($\sim 150-200$ eV)
remains. Furthermore, the \ginga and \asca results (Smith \& Done 1996,
Turner et al. 1997) showed a cold iron line with EW $\sim 50$ and $\sim
70$ eV respectively (these estimates had large errors associated with
them). The continuum flux was $\sim 4 \times 10^{-11}$ erg s $^{-1}$
cm$^{-2}$ in both cases. Despite the large errors, for this source flux
level, the equivalent widths are in agreement with these expected
according to EW vs flux plot for this source in Fig~7. Guainazzi et al.
(1998) report the same EW when the source was $\sim 4$ times fainter.
Their estimate appears to be inconsistent with our results, and suggests
that the line follows the continuum on time scales of $\sim$ a year. The
\xte observations were made within a week. One explanation could then be
that the line is mainly produced by the torus which is located at a
distance larger than $\sim$ a light week and smaller than $\sim$ a light
year.

ESO103-G35 was observed three times by \asca (Forster et al. 1999). In all
cases, the source flux was lower than during the \xte observations. The EW
of the iron line was higher ($\sim 400-200$ eV), in agreement with the
trend shown in Fig. \ref{ew}. 
As in the case of NGC7172, the large EW observed by
\asca requires an extra source of line photons apart from the torus and
the accretion disk.  Forster et al. (1999) suggest various possibilities
in
order to explain this. The fact that the iron line appears broad in the
\xte observations is a direct indication that a part of the line is
produced by fluorescence from the accretion disk. In fact, ESO103-G35 is
the only source were a model with a variable iron line (model D') improves
significantly the goodness fit when compared to model A'. This indication
of line variability supports the idea that the spectral variability in
this object is due to a variable reflection component (model E'). Such a
model could also explain naturally the fact that $HR2$ in the case of
ESO103-G35 is variable while $HR1$ remains constant. The possible
detection of line and reflection component variability in this object only
indicates that the orientation of the central source may not be the same
in Seyfert 2 galaxies (despite the predictions of the unification models
in their simplest form). Our results are consistent with the idea that the
central source in ESO103-G35 is seen close to face on and at a larger
inclination in the other objects. In the latter case, the contribution
from the reflection component and the disk iron line emission will be
supressed and it will be more difficult to detect any variations
associated with them.

The main result from the \xte observations of the four Seyfert 2 galaxies
ESO103-G35, IC5063, NGC4507 and NGC7172 is that they all show significant
spectral variability in the sense that the spectrum steepens 
 with increasing flux. This is a model independent result and is based on
the analysis of the hardness ratios (see Section 4.1, Fig.~2). The origin
of this variability may not be the same for all of them. In NGC4507 and
NGC7172 we observe flux related $N_{H}$ and $\Gamma$ variations
respectively. In ESO103-G35 the spectral variability could be due either
to photon index variations (like NGC7172) or to reflection component
changes. The results for IC5063 are puzzling. The stability of the iron
line flux does not support reflection component variations as the reason
of the observed spectral variations. On the other hand, the photon index
variations are not well correlated with the continuum flux. Since this
object is a radio galaxy, perhaps the normal ``Seyfert - type" X-ray
source in this case is ``contaminated" by emission from the non-thermal
source which is responsible for the radio emission.

The differences in the origin of the spectral variability observed in
these objects suggest that the orientation of the nucleus and/or the
distribution of the absorbing matterial around the nucleus is not the same
in all Seyfert 2 galaxies. We believe that future monitoring observations
of these objects with XMM mission which has both a large effective area
and good spectral resolution are necessary in order to clarify the issue of
spectral variability in Seyfert 2 galaxies and provide stringent
constraints on the geometry of the nucleus and the location of the
cirumnuclear matter.

\section{Acknowledgements}
We are grateful to D.A. Smith for numerous conversations and suggestions
regarding the \xte background.  This research has made use of data
obtained from the High Energy Astrophysics Science Archive Research Center
(HEASARC), provided by NASA's Goddard Space Flight Center.


\begin{table*}
\caption{Power-law and Gaussian line fits for ESO103-G35}
\label{eso103_poga}
\begin{tabular}{cccccc}
Model  &  $\rm N_H^5$ &  $\Gamma$ & Energy (keV) & $\sigma$ (keV) & \x2/dof \\
\hline
A$^1$  & $14.1^{+0.5}_{-0.4}$ &  $1.57^{+0.02}_{-0.03}$
  & $5.93^{+0.01}_{-0.01}$ & $0.69^{+0.05}_{-0.04}$ & 791.3/730 \\
B$^2$ & $12.6^{+0.6}_{-0.6}-16.0^{+1.0}_{-1.0}$ &  $1.55^{+0.03}_{-0.03}$
  & $5.91^{+0.04}_{-0.04}$ & $0.72^{+0.05}_{-0.05}$ & 684.3/716 \\
C$^3$ & $13.9^{+0.4}_{-0.4}$&  $1.47^{+0.17}_{-0.10}-1.61^{+0.04}_{-0.04}$
  & $5.90^{+0.04}_{-0.04}$ & $0.73^{+0.05}_{-0.05}$ & 629.4/716 \\
D$^4$ & $13.8^{+0.5}_{-0.5}$&  $1.54^{+0.02}_{-0.02}$
  & $5.90^{+0.05}_{-0.05}$ & 0.1 & 723.7/716 \\
\hline
\end{tabular}

Notes: $^1$ $N_H$, $\Gamma$ tied; $^2$ $N_H$  untied,
 $\Gamma$ tied; $^3$ $N_H$ tied, $\Gamma$ untied; \\ $^4$
 gaussian line normalizations untied ; $^5$ in units of
 $10^{22}$ \cunits
\end{table*}

\begin{table*}
\caption{Power-law and Gaussian line fits for IC5063}
\label{ic5063_poga}
\begin{tabular}{cccccc}
Model  &  $\rm N_H^5$ &  $\Gamma$ & Energy (keV) & $\sigma$ (keV) &
\x2/dof \\
\hline
A$^1$  & $19.2^{+1.7}_{-1.7}$&  $1.57^{+0.07}_{-0.07}$
  & $6.11^{+0.08}_{-0.09}$ & $0.37^{+0.13}_{-0.17}$ & 310.0/240 \\
B$^2$ & $15.0^{+4.0}_{-4.0}-28.0^{+2.0}_{-3.0}$ &  $1.52^{+0.10}_{-0.09}$
  & $6.02^{+0.10}_{-0.03}$ & $0.51^{+0.14}_{-0.06}$ & 191.9/236 \\
C$^3$ & $17.9^{+1.6}_{-1.6}$&  $1.15^{+0.10}_{-0.13}-
1.70^{+0.20}_{-0.19}$
  & $6.05^{+0.09}_{-0.10}$ & $0.50^{+0.15}_{-0.14}$ & 177.2/236 \\
D$^4$ & $17.8^{+2.5}_{-1.3}$&  $1.47^{+0.15}_{-0.06}$
  & $6.04^{+0.13}_{-0.14}$ & $0.53^{+0.09}_{-0.27}$ & 302.6/236 \\
\hline
\end{tabular}

Notes: $^1$ $N_H$, $\Gamma$ tied; $^2$ $N_H$  untied,
 $\Gamma$ tied; $^3$ $N_H$ tied, $\Gamma$ untied; \\ $^4$
 gaussian line normalizations untied ; $^5$ in units of
 $10^{22}$ \cunits
\end{table*}

\begin{table*}
\caption{Power-law and Gaussian line fits for NGC4507}
\label{ngc4507_poga}
\begin{tabular}{cccccc}
Model  &  $\rm N_H^5$ &  $\Gamma$ & Energy (keV) & $\sigma$ (keV) &
\x2/dof \\
\hline
A$^1$  & $45.0^{+1.5}_{-1.5}$&  $1.40^{+0.05}_{-0.01}$ 
  & $6.09^{+0.04}_{-0.06}$ & $0.33^{+0.04}_{-0.06}$ & 887.2/807 \\
B$^2$ & $40.3^{+1.0}_{-1.0}-50.9^{+2.0}_{-3.0}$ &  $1.40^{+0.05}_{-0.03}$
  & $6.03^{+0.05}_{-0.15}$ & $0.42^{+0.05}_{-0.05}$ & 742.5/794 \\
C$^3$ & $45.2^{+2}_{-2}$&  $1.27^{+0.06}_{-0.01}- 1.53^{+0.05}_{-0.04}$
  & $6.05^{+0.05}_{-0.04}$ & $0.40^{+0.04}_{-0.03}$ & 763.7/794 \\
D$^4$ & $46.7^{+1}_{-1}$&  $1.42^{+0.03}_{-0.03}$
  & $5.96^{+0.03}_{-0.03}$ & $0.52^{+0.04}_{-0.04}$ & 815.8/794 \\
\hline
\end{tabular}

Notes: $^1$ $N_H$, $\Gamma$ tied; $^2$ $N_H$  untied,
 $\Gamma$ tied; $^3$ $N_H$ tied, $\Gamma$ untied; \\ $^4$
 gaussian line normalizations untied ; $^5$ in units of
 $10^{22}$ \cunits
\end{table*}

\begin{table*}
\caption{Power-law and Gaussian line fits for NGC7172}
\label{ngc7172_poga}
\begin{tabular}{cccccc}
Model  &  $\rm N_H^5$ &  $\Gamma$ & Energy (keV) & $\sigma$ (keV) &
\x2/dof \\
\hline
A$^1$  & $11.5^{+0.3}_{-0.8}$ &  $1.81^{+0.03}_{-0.05}$
  & $6.19^{+0.09}_{-0.09}$ & 0.1 & 304.2/437 \\
B$^2$ & $11.0^{+1.0}_{-1.0}-14.0^{+1.0}_{-1.0}$&  $1.81^{+0.06}_{-0.05}$
  & $6.18^{+0.08}_{-0.10}$ & 0.1 & 274.6/429 \\
C$^3$ & $11.5^{+0.7}_{-0.8}$&  $1.66^{+0.14}_{-0.16}-1.96^{+0.03}_{-0.08}$
  & $6.19^{+0.09}_{-0.09}$ & 0.1 & 274.7/429 \\
D$^4$ & $11.3^{+0.5}_{-0.5}$&  $1.80^{+0.01}_{-0.04}$
  & $6.17^{+0.08}_{-0.08}$ & 0.1 & 299.8/429 \\
\hline
\end{tabular}

Notes: $^1$ $N_H$, $\Gamma$ tied; $^2$ $N_H$  untied,
 $\Gamma$ tied; $^3$ $N_H$ tied, $\Gamma$ untied; \\ 
 $^4$ gaussian line normalizations untied ; $^5$ in units of
 $10^{22}$ \cunits
\end{table*}


\begin{table*}
\caption{Reflection component fits for ESO103-G35}
\label{eso103_pogapex}
\begin{tabular}{cccccc}
Model  &  $\rm N_H^1$ &  $\Gamma$ & Energy (keV) & $\sigma$ (keV) &
\x2/dof \\
\hline
A'$^2$  & $17.2^{+0.3}_{-0.2}$ &  1.9
  & $6.07^{+0.03}_{-0.03}$ & $0.5^{+0.10}_{-0.10}$  & 627.5/730 \\
B'$^3$ & $15.9^{+1.0}_{-1.0}-19.0^{+1.0}_{-1.0}$&  1.9
  & $6.06^{+0.05}_{-0.05}$ & $0.5^{+0.09}_{-0.03}$  & 556.2/716 \\
C'$^4$ & $17.5 ^{+0.2}_{-0.2}$&
$1.84^{+0.08}_{-0.08}-2.02^{+0.05}_{-0.05}$
  & $6.07^{+0.06}_{-0.06} $ & $0.47^{+0.08}_{-0.04}$   & 523.5/715    \\
D'$^5$ & $17.2^{+0.2}_{-0.2}$&  1.9
  & $6.06^{+0.05}_{-0.05}$ & $0.51^{+0.08}_{-0.04}$  & 604.7/716 \\
E'$^6$ & $17.3^{+0.2}_{-0.2}$ & 1.9
  & $6.06^{+0.04}_{-0.05}$ & $0.51^{+0.08}_{-0.04}$ & 525.8/716 \\
\hline
\end{tabular}
 
Notes: $^1$ in units of $10^{22}$ \cunits;
  $^2$ $N_H$, $\Gamma$ tied; $^3$ $N_H$  untied,
 $\Gamma$ tied; $^4$ $N_H$ tied, $\Gamma$ untied; \\
 $^5$ gaussian line normalizations untied ; $^6$
 reflection component normalisation untied
\end{table*}

\begin{table*}
\caption{Reflection component fits for IC5063}
\label{ic5063_pogapex}
\begin{tabular}{cccccc}
Model  &  $\rm N_H^1$ &  $\Gamma$ & Energy (keV) & $\sigma$ (keV) &
\x2/dof \\
\hline
A'$^2$  & $21.9^{+1.0}_{-1.0}$ &  1.9
  & $6.18^{+0.10}_{-0.10}$ & 0.1  & 269.3/241 \\
B'$^3$ & $18.4^{+4.0}_{-4.0}-31.0^{+2.5}_{-3.5}$&  1.9
  & $6.17^{+0.05}_{-0.05}$ & 0.1  & 186.8/237 \\
C'$^4$ & $22.7^{+0.8}_{-0.8}$&
$1.57^{+0.10}_{-0.10}-2.20^{+0.15}_{-0.15}$
  & $6.18^{+0.05}_{-0.05} $ & 0.1  & 181.8/236    \\
D'$^5$ & $21.9^{+1.0}_{-1.0}$&  1.9
  & $6.20^{+0.07}_{-0.09}$ & 0.1  & 264.3/237 \\
E'$^6$ & $22.3^{+1.0}_{-1.0}$ & 1.9
  & $6.18^{+0.10}_{-0.10}$ & 0.1 & 176.7/237 \\
\hline
\end{tabular}

Notes: $^1$ in units of $10^{22}$ \cunits;
  $^2$ $N_H$, $\Gamma$ tied; $^3$ $N_H$  untied,
 $\Gamma$ tied; $^4$ $N_H$ tied, $\Gamma$ untied; \\
 $^5$ gaussian line normalizations untied ; $^6$
reflection component normalisation untied 
\end{table*}  

\begin{table*}
\caption{Reflection component fits for NGC4507}
\label{ngc4507_pogapex}
\begin{tabular}{cccccc}
Model  &  $\rm N_H^1$ &  $\Gamma$ & Energy (keV) & $\sigma$ (keV) &
\x2/dof \\
\hline
A'$^2$  & $50.4^{+1.1}_{-0.6}$ &  1.9
  & $6.14^{+0.03}_{-0.04}$ & 0.1  & 737.2/807 \\
B'$^3$ & $47.9^{+0.8}_{-0.8}-53.4^{+1.2}_{-1.2}$&  1.9
  & $6.13^{+0.04}_{-0.04}$ & 0.1  & 701.5/795 \\
C'$^4$ & $51.9^{+1.0}_{-1.0}$&
$1.90^{+0.05}_{-0.05}-2.01^{+0.07}_{-0.07}$
  & $6.13^{+0.04}_{-0.04} $ & 0.1  & 700.7/794   \\
D'$^5$ & $50.8^{+1.0}_{-1.0}$&  1.9
  & $6.13^{+0.04}_{-0.04}$ & 0.1  & 732.0/795 \\
E'$^6$ & $50.8^{+1.0}_{-1.0}$ & 1.9
  & $6.14^{+0.03}_{-0.04}$ & 0.1 & 717.8/795 \\
\hline
\end{tabular}

Notes: $^1$ in units of $10^{22}$ \cunits;
  $^2$ $N_H$, $\Gamma$ tied; $^3$ $N_H$  untied,
 $\Gamma$ tied; $^4$ $N_H$ tied, $\Gamma$ untied; \\
 $^5$ gaussian line normalizations untied ; $^6$
 reflection component normalisation untied
\end{table*}  

\begin{table*}
\caption{Reflection component fits for NGC7172}
\label{ngc7172_pogapex}
\begin{tabular}{cccccc}
Model  &  $\rm N_H^1$ &  $\Gamma$ & Energy (keV) & $\sigma$ (keV) &
\x2/dof \\
\hline
A'$^2$  & $12.2^{+0.4}_{-0.4}$ &  1.9
  & $6.21^{+0.09}_{-0.09}$ & 0.1 & 299.7/436 \\ 
B'$^3$ & $9.9^{+1.0}_{-1.0}-14.0^{+3.0}_{-3.0}$&  1.9
  & $6.21^{+0.10}_{-0.10}$ & 0.1 & 273.5/428 \\
C'$^4$ & $11.7^{+0.8}_{-0.4}$&
$1.74^{+0.15}_{-0.22}-1.99^{+0.08}_{-0.07}$
  & $6.20^{+0.10}_{-0.10}$ & 0.1 & 274.1/428 \\ 
D'$^5$ & $12.2^{+0.4}_{-0.2}$&  1.9
  & $6.20^{+0.10}_{-0.10}$ & 0.1 & 297.1/428 \\
E'$^6$ & $12.1^{+0.5}_{-0.5}$ & 1.9
  & $6.20^{+0.10}_{-0.08}$ & 0.1 & 280.7/428 \\
\hline
\end{tabular}
  
Notes: $^1$ in units of $10^{22}$ \cunits;
  $^2$ $N_H$, $\Gamma$ tied; $^3$ $N_H$  untied,
 $\Gamma$ tied; $^4$ $N_H$ tied, $\Gamma$ untied; \\
 $^5$ gaussian line normalizations untied ; $^6$
 reflection component normalisation untied
\end{table*}

\end{document}